\documentclass[aps,prl,reprint,superscriptaddress, floatfix]{revtex4-2}

\usepackage{graphicx}
\usepackage{dcolumn}
\usepackage{bm}
\usepackage{placeins}
\usepackage{color}
\usepackage{xcolor}
\usepackage{epstopdf}
\usepackage{gensymb}
\usepackage{ulem}
\usepackage{float}
\usepackage{xr}
\usepackage{xr-hyper}
\usepackage{mathrsfs}
\usepackage{amsmath}
\usepackage{amsthm}
\usepackage{enumerate}
\usepackage{soul}
\usepackage{amssymb}
\usepackage{units}
\usepackage{todonotes}
\usepackage{verbatim}
\usepackage{hyperref}
\usepackage{sidecap}
\usepackage{wrapfig}
\usepackage{caption}
\usepackage{cleveref}
\usepackage{dsfont}
\usepackage{units}
\usepackage{lipsum}
\usepackage{comment}
\usepackage{bm}
\usepackage{textcomp, gensymb}
\begin{document}
\preprint{APS/123-QED}
\title{Quantum Computing for Solid Mechanics and Structural Engineering\\ -- a Demonstration with Variational Quantum Eigensolver}

\author{Yunya Liu}
\affiliation{Department of Mechanical Engineering, University of Utah, Salt Lake City, UT, USA}

\author{Jiakun Liu}
\affiliation{Department of Mechanical Engineering and Applied Mechanics, University of Pennsylvania, Philadelphia, PA, USA}

\author{Jordan R. Raney}
\affiliation{Department of Mechanical Engineering and Applied Mechanics, University of Pennsylvania, Philadelphia, PA, USA}

\author{Pai Wang}
\affiliation{Department of Mechanical Engineering, University of Utah, Salt Lake City, UT, USA}

\date{\today}
             
\begin{abstract}
Variational quantum algorithms exploit the features of superposition and entanglement to optimize a cost function efficiently by manipulating the quantum states. They are suitable for noisy intermediate-scale quantum (NISQ) computers that recently became accessible to the worldwide research community. 
Here, we implement and demonstrate the numerical processes on the 5-qubit and 7-qubit quantum processors on the IBM Qiskit Runtime platform. 
We combine the commercial finite-element-method (FEM) software \textsc{abaqus} with the implementation of \textit{Variational Quantum Eigensolver} (VQE) to establish an integrated pipeline. Three examples are used to investigate the performance: a hexagonal truss, a Timoshenko beam, and a plane-strain continuum. We conduct parametric studies on the convergence of fundamental natural frequency estimation using this hybrid quantum-classical approach. Our findings can be extended to problems with many more degrees of freedom when quantum computers with hundreds of qubits become available in the near future.
\end{abstract}

\maketitle

\section{Introduction}
Variational quantum algorithms (VQAs)~\cite{cerezo2021variational} can solve problems 
in optimization~\cite{zhu2022adaptive}, machine learning~\cite{sajjan2022quantum, divya2021quantum}, physics~\cite{weimer2021simulation}, chemistry~\cite{mcardle2020quantum}, material sciences~\cite{bauer2020quantum}, and cryptography~\cite{bennett2020quantum}. Due to the unique features of entanglement and superposition, quantum computers can leverage VQAs to tackle problems efficiently and accurately while bypassing the limitation of memory allocation and computational complexity. VQAs achieve the goal of finding the solution by integrating classical optimizers with a quantum circuit. The quantum part here is designed to prepare quantum states and their measurements thereafter while the classical part is used to tune the quantum circuit parameters. 
The potential advantages of VQAs lie in the scaling law with respect to degrees of freedom (DOFs) in the mathematical model of problems: A particular quantum state represented by $N$ qubits 
can encode the information of $2^N$ DOFs. This translates to an exponential scaling that surpasses any possible classical computing process, thereby facilitating a significant speed-up as compared to traditional solvers. 

Specialized algorithms within VQAs suitable for near-term noisy intermediate-scale quantum (NISQ) computers were recently demonstrated in analyzing electronic structures~\cite{huang2022simulating}, molecular spectra~\cite{barone2021computational}, fluid flows~\cite{bharadwaj2020quantum, demirdjian2022variational}, heat transfer~\cite{liu2022application, lu2020quantum} 
, as well as general algebraic~\cite{bravo2019variational, xu2021variational} and differential systems~\cite{lubasch2020variational, fang2023time, krovi2023improved}.
While these recent studies exemplify the vast potential of quantum computing and its implementation, the research community has yet to showcase an integrated pipeline that unifies VQAs with the finite-element method (FEM), which is a robust and widely-used technique across many disciplines. 
Motivated by this gap, we aim to investigate strategies for deploying quantum solvers to solve eigenvalue problems that are ubiquitous in mechanical systems. 
Focusing specifically on vibration analyses, the precise identification of the fundamental natural frequency in structures is critical in engineering practice. The importance is underscored by the resonant phenomenon where a small oscillatory perturbation can provoke a disproportionally large response.


In this Letter, we combine the FEM capability of the commercial software package \textsc{abaqus} together with the \textit{Variational Quantum Eigensolver} (VQE)~\cite{peruzzo2014variational, ferguson2021measurement, tilly2022variational} on Qiskit quantum computing platform to implement an integrative FEM-VQE pipeline aiming at finding the fundamental natural frequency of different structures. We demonstrate and analyze three example cases: 
(I) hexagonal truss, (II) Timoshenko beam, and (III) plane-strain continuum. We first test our hybrid quantum-classical algorithm on a simulator backend and then on quantum processing units (QPUs) with 3$\sim$7 qubits. Results from classical solvers are used as benchmarks to quantify the errors. We perform a series of parametric studies on the key factors in the implementation. In addition, we also discuss current limitations and potential future directions. 

\begin{figure*}[!ht]
    \centering
    \includegraphics[width=1.0\textwidth]{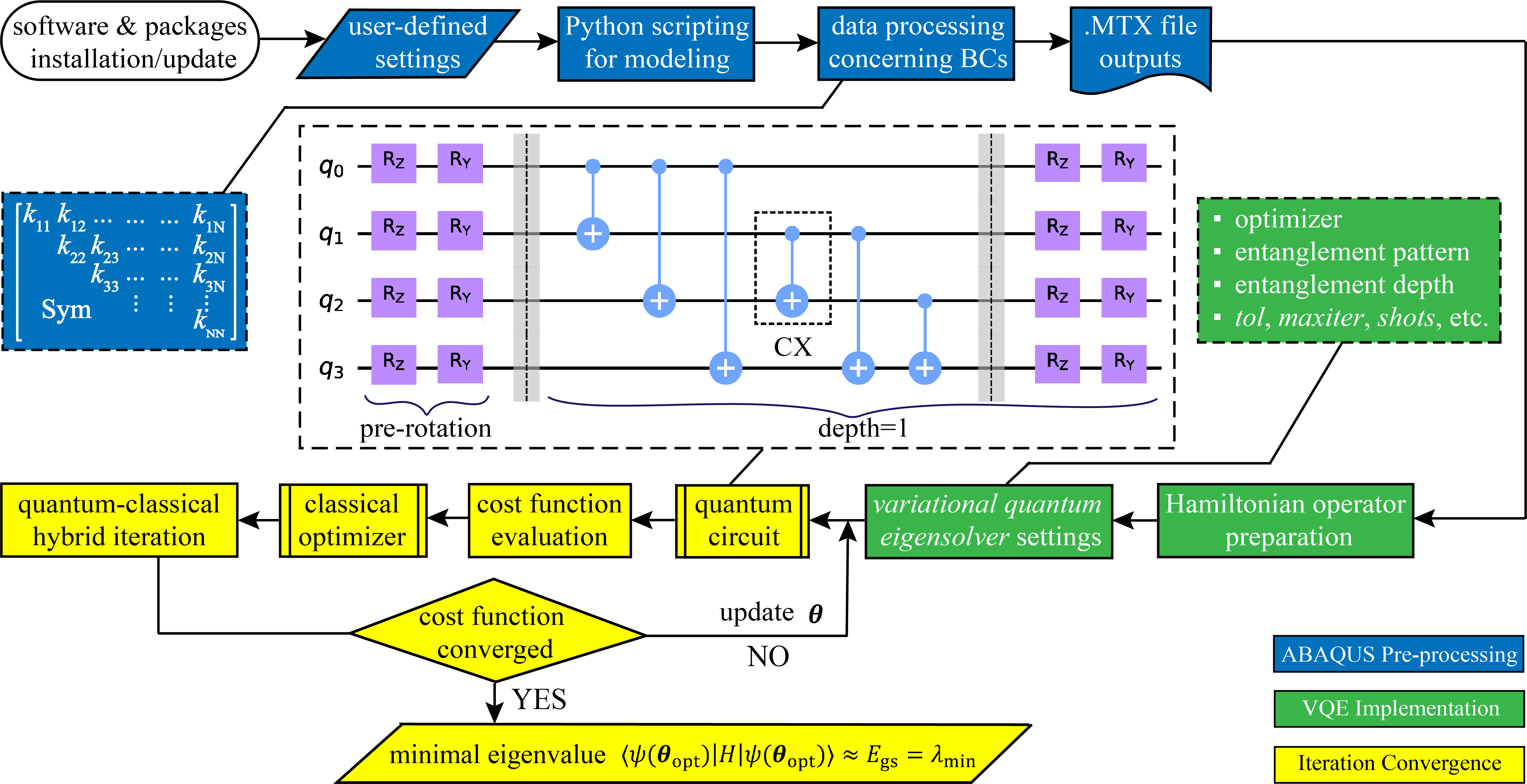}
    \caption{\small Pipeline of \textit{Variational Quantum Eigensolver} algorithm for mechanics problems. The blue-colored steps refer to \textsc{abaqus} Pre-processing, those in green to VQE Implementation, and the remaining in yellow to Iteration Convergence. 
    An example stiffness matrix is shown in the left dashed-frame inset. 
    A particular instance of the ``hardware-efficient ansatz"~\cite{kandala2017hardware, you2021exploring} 
    is depicted in the central dashed-frame inset, 
    where the entanglement pattern with 6 CX gates and depth = 1 is shown as an example. 
    Additional examples are presented in \textit{Supplemental Material}~\cite{SI}. The right dashed-frame inset lists the key user-defined parameters. 
    }
    \label{fig:img1}
\end{figure*}

\section{Numerical Implementation}
The quantum computing research community is rapidly expanding, and a variety of toolsets with different architectures are available (in chronological order of release dates): Forest SDK/PyQuil~\cite{smith2016practical, Karalekas_2020}, Microsoft Quantum Development Kit~\cite{svore2018q}, ProjectQ~\cite{steiger2018projectq}, PennyLane~\cite{bergholm2018pennylane}, Amazon Braket~\cite{Braket}, IBM Qiskit~\cite{aleksandrowicz2019qiskit}, Cirq/TensorFlow Quantum~\cite{omole2020cirq, broughton2020tensorflow}, and Bosonic Qiskit~\cite{stavenger2022bosonic}. 
Here, we employ the IBM Qiskit platform, primarily due to its maturity and extensive community support.

Fig.\,\ref{fig:img1} depicts the steps to realize the integrative quantum-classical pipeline. Our approach involves the following components:
\begin{itemize}
  \item \textsc{abaqus/cae} and \textsc{abaqus/standard} 
  \item Qiskit quantum hardware and simulators 
  \item NumPy and Matplotlib
\end{itemize}
There are three major parts of this hybrid framework:\\
\textbf{(A) \textsc{ABAQUS} pre-processing} is shown as the blue-colored steps in Fig.\,\ref{fig:img1}.
Within the \textsc{abaqus/cae}  environment, a scripting approach converts the input of user-defined problem settings to the \textsc{.mtx} file outputs. The stiffness and mass matrices are the outcome of data processing. 

\textbf{(B) VQE implementation} is shown as green-colored steps in Fig.\,\ref{fig:img1}. It consists of processes mapping the Hamiltonian operator to VQE settings, which include the choice of the classical optimizer, entanglement pattern, and entanglement depth. Fig.\,\ref{fig:img1} shows an example of a quantum circuit with entanglement pattern CX and entanglement depth = 1 applied on four qubits. In addition, there are several additional tunable parameters: 1) optimizer-dependent tolerance; 2) quantum processor-dependent shots, transpile optimization, coherence time, statistical quantum state measurement, and error mitigation techniques; as well as 3) maximum number of iterations which depends on both the optimizer and quantum processor.

\textbf{(C) Iterative convergence} includes yellow-colored steps shown in Fig.\,\ref{fig:img1}. They span from quantum circuit design to the decision-making of cost function convergence, iterating until the minimal eigenvalue estimate is obtained. 
The convergence criterion, represented by the tolerance parameter, ensures convergence when the absolute difference between evaluations of two consecutive cost functions is smaller than the tolerance parameter, indicating proximity to the optimal solution of VQE implementation.

\subsection{(A) \textsc{ABAQUS} Pre-processing}
We develop pre-processing commands to generate a FEM model with a targeted number of free DOFs. 
The overall goal is to find a suitable combination of the total number of degrees of freedom $n^\textrm{all}_\textrm{DOF}$ and the prescribed (fixed) degrees of freedom $n^\textrm{fixed}_\textrm{DOF}$ in the model such that $2^N = n^\textrm{all}_\textrm{DOF} - n^\textrm{fixed}_\textrm{DOF}$ where $N$ is an integer. 
The general procedures are summarized as follows:\\
(1) Employ pre-processing Python script in the \textsc{abaqus/cae} environment to create an initial model with corresponding geometry, material properties, and boundary conditions. \\
(2) Mesh the model with appropriate element types: 
T2D2 for the hexagonal truss, B21 for the Timoshenko beam, and CPE3/CPE4 for the plane-strain continuum. 
(3) Adjust and modify the mesh of the model to reach the targeted number of free DOFs determined by $N$. For the beam system, a proper element size can be directly calculated for a given $N$, whereas for truss and continuum 
systems, we perform an iterative approach to adjust the assigned global element size (with subsequent re-meshing). More details are presented in the \textit{Supplemental Material}~\cite{SI}. \\
(4) Once the targeted free DOFs has been reached, export mass matrix $M$ and stiffness matrix $K$ from \textsc{abaqus}. These matrices, by default, will include all DOFs as a result of the internal processing logistics. \\
(5) Execute another Python script 
to read the exported mass and stiffness matrices from \textsc{.mtx} files. Then, apply partitioning to get the 
matrices with only free DOFs by filtering out columns and rows associated with boundary conditions. 
\subsection{(B) VQE Implementation}
Quantum computing's potency, particularly in the domain of NISQ computers, extends beyond quantum mechanical systems. The physical interpretation of vibration must be adapted to align with the noise-tolerant and shallow entanglement depth of NISQ devices. Here, we demonstrate two types of implementations: a noise-free simulator and quantum processors~\cite{IBM_Newsroom}. 
The underpinning concept is centered on the VQE, which is tasked with identifying the ground state energy of a Hamiltonian. This gives the fundamental natural frequency of a mechanical system~\cite{peruzzo2014variational, liu2019variational, yordanov2021qubit}.

Generally, a $2^{N}\times2^{N}$ square matrix needs $N$ qubits to encode it as a quantum Hamiltonian. From the \textsc{.mtx} files exported from \textsc{abaqus}, we prepare and decompose our Hamiltonian as 
\begin{equation}
    H = M^{-1}K = \sum c_{l} P_{l}
    \label{eq:decomp}
\end{equation}
where $P_{l} \in \{I, X, Y, Z \}^{\otimes N}$ represents a multi-qubit ($N$-qubit) Pauli operator, and $c_{l}$'s are coefficients of decomposition.


The accuracy of the VQE depends on three key factors including classical optimizer~\cite{moll2018quantum, bonet2023performance}, entanglement pattern~\cite{liu2019variational, zhang2022quantum}, and entanglement depth~\cite{malvetti2021quantum, gleinig2021efficient}. In this study, we test all of the following.

- Classical optimizers: 
\begin{itemize}
  \item Simultaneous Perturbation Stochastic Approximation (SPSA) 
  - a non-gradient-based method
  \item Constrained Optimization by Linear Approximation (COBYLA) 
  - a non-gradient-based method
  \item Sequential Least Squares Programming (SLSQP) 
  - a gradient-based method
  \item Limited memory Broyden Fletcher Goldfarb Shanno Bound (L-BFGS-B) 
  - a gradient-based method
\end{itemize}
- Entanglement patterns:
\begin{itemize}
  \item Controlled Not (CNOT or CX) gate
  \item Controlled Z (CZ) gate 
  \item Controlled Rotation X (CRX) gate 
\end{itemize}
- Entanglement depths: 
\begin{itemize}
\item An integer (1$\sim$10) specifying the quantum circuit 
\end{itemize}

Together, these components constitute a hardware-efficient ansatz (i.e., a parameterized quantum circuit) 
in the variational form of the ``EfficientSU2" class~\cite{peruzzo2014variational}. 
As a specific example, the central inset in Fig.\,\ref{fig:img1} illustrates a particular instance of the ansatz comprising $N=4$ qubits, utilizing 
single-qubit quantum gates $R_{z}$ and $R_{y}$, and 
adopting depth $= 1$ with entanglement pattern CX. 
First, quantum states are prepared for the qubits, and their initial joint state can be written as~\cite{CMUlecture}
\begin{equation}
    | \psi \rangle ^{\otimes N}  
    = | \psi_{N-1} \rangle \otimes ...\otimes | \psi_1 \rangle  \otimes | \psi_{0} \rangle
     = \bigotimes_{j=0}^{N-1} | \psi_j \rangle 
    \label{eq:initial}
\end{equation}
where $\otimes$ denotes the tensor product between quantum states.
They then go through two layers or ``slices" of pre-rotations~\cite{Qiskit, wong2022introduction} represented by the pre-operator $U_\textrm{pre}$, and this results in
\begin{equation}
\begin{split}
    &\left[\bigotimes_{j=0}^{N-1} R_y(\theta_{2,j})\right]
    \left[\bigotimes_{j=0}^{N-1} R_z(\theta_{1,j})\right]
    \left[\bigotimes_{j=0}^{N-1} | \psi_j \rangle \right]\\
    = 
    &\bigotimes_{j=0}^{N-1} \Big[ R_y(\theta_{2,j}) R_z(\theta_{1,j}) |\psi_j\rangle \Big] 
    = U_\textrm{pre}  |\psi\rangle^{\otimes N}.
    \label{eq:pre-rotation}
\end{split}
\end{equation}
Next, the states are fully entangled together by the entanglement operator consisting of two-qubit (non-parameterized) CX gates,
\begin{equation}
   U_\textrm{ent} = 
   \prod_{j_1=0}^{N-2} \;
   \prod_{j_2=j_1+1}^{N-1}
   \textrm{CX}_{j_1,j_2},
\label{eq:cx}
\end{equation}
which is followed by another two ``slices" of rotations,
\begin{equation}
   U_\textrm{rot} = 
   \left[\bigotimes_{j=0}^{N-1} R_y(\theta_{4,j})\right]
   \left[\bigotimes_{j=0}^{N-1} R_z(\theta_{3,j})\right],
\label{eq:rot}
\end{equation}
where 
$\boldsymbol{\theta} = \{ \theta_{s,j} \}$ is a set of variational parameters, which control all the single-qubit gates (i.e., $R_{y}$ and $R_{z}$). Here, $s = 1,..,4$ indicates the ``slice" of the single-qubit gates, and $j, j_1, j_2 \in \{0, 1, ..., N-1\}$ are indices of qubits. The combined effect of Eqs.\,\eqref{eq:cx} and \eqref{eq:rot} constitutes one ``depth", and the total operator of the parameterized variational ansatz is 
\begin{equation}
   U(\boldsymbol{\theta}) = [U_\textrm{rot}] [U_\textrm{ent}] [U_\textrm{pre}].
\label{eq:ansatz}
\end{equation}
Alternative implementations of other entanglement patterns and expressions are presented in \textit{Supplemental Material}~\cite{SI}.

\subsection{(C) Iterative Convergence}
The quantum-classical hybrid iteration process successively alters the quantum circuit parameters to minimize the cost function, which is defined as:
 \begin{equation}
    C(\boldsymbol{\theta}) = \langle \psi(\boldsymbol{\theta}) | H | \psi(\boldsymbol{\theta}) \rangle,
    \label{eq:cost}
    \end{equation}
where $| \psi(\boldsymbol{\theta}) \rangle$ is the output quantum state determined by parameters $\boldsymbol{\theta} = \{ \theta_{s,j} \}$ shown in Eqs.\,\eqref{eq:pre-rotation} and \eqref{eq:rot}.
This process iterates until it reaches either the user-defined tolerance $tol$ or the maximum number of iterations $maxiter$. In the following, we list all the pre-defined parameters, which we use to generate all data presented in this Letter.
\begin{itemize}
    \item  $tol$ represents the convergence criterion, which is used against the absolute approximate error:
    \begin{equation}
        | E_j - E_{j-1}| < tol,     
        \label{eq:tol}
    \end{equation}
    where 
    $E_{j}$ is the ground state energy estimate of the $j$-th iteration. 
    \item $maxiter$ denotes the maximum number of iterations that an optimizer is allowed to go through. 
    \item $shots$ is the number of times the quantum circuit is executed for each evaluation of the cost function, determining the statistical accuracy of $E_\textrm{j}$. 
\end{itemize}
In this study, the noise-free simulator uses $shots=10^5$ and $maxiter=10^5$, while QPUs employ $shots=2 \times 10^4$ and $maxiter=100$, which are the largest possible on the quantum hardware. 
In all cases, we set $tol=10^{-4}$.\\

The hybrid algorithm iteratively updates the quantum-circuit parameters $\boldsymbol{\theta}$ using protocols encoded in the classical optimizer. 
Then a new quantum state is generated as
\begin{equation}
     |\psi(\boldsymbol{\theta}) \rangle = U(\boldsymbol{\theta})| \psi \rangle ^{\otimes N},
    \label{eq:qstate}
\end{equation}
where $| \psi \rangle ^{\otimes N}$ denotes the initial quantum states defined in Eq.\,\eqref{eq:initial}. 
The new state $| \psi(\boldsymbol{\theta}) \rangle$ is then used to evaluate the cost function defined in Eq.\,\eqref{eq:cost}.
When the threshold $tol$ is reached by iterations, the optimization yields the final expectation value (i.e., minimized cost function) as
\begin{equation}
     C(\boldsymbol{\theta}_\textrm{opt}) = \langle \psi(\boldsymbol{\theta}_\textrm{opt}) | H | \psi(\boldsymbol{\theta}_\textrm{opt}) \rangle \approx E_\textrm{gs},
    \label{eq:cos_opt}
\end{equation}
where $\boldsymbol{\theta}_\textrm{opt}$ and $E_\textrm{gs}$ denote the optimal set of parameters and the quantum ground state energy, respectively.
Upon convergence, we obtain
\begin{equation}
   \lambda_\textrm{min} \approx E_\textrm{gs} \quad \text{and} \quad |\psi_\textrm{min} \rangle \approx |\psi(\boldsymbol{\theta}_\textrm{opt}) \rangle, 
    \label{eq:final_min}
\end{equation}
where $\lambda_\textrm{min}$ is the minimum eigenvalue, 
and $|\psi_\textrm{min} \rangle $ is the quantum state corresponding to $\lambda_\textrm{min}$. 

\begin{figure}[t!]
\centering
\includegraphics[width=1.0\linewidth]{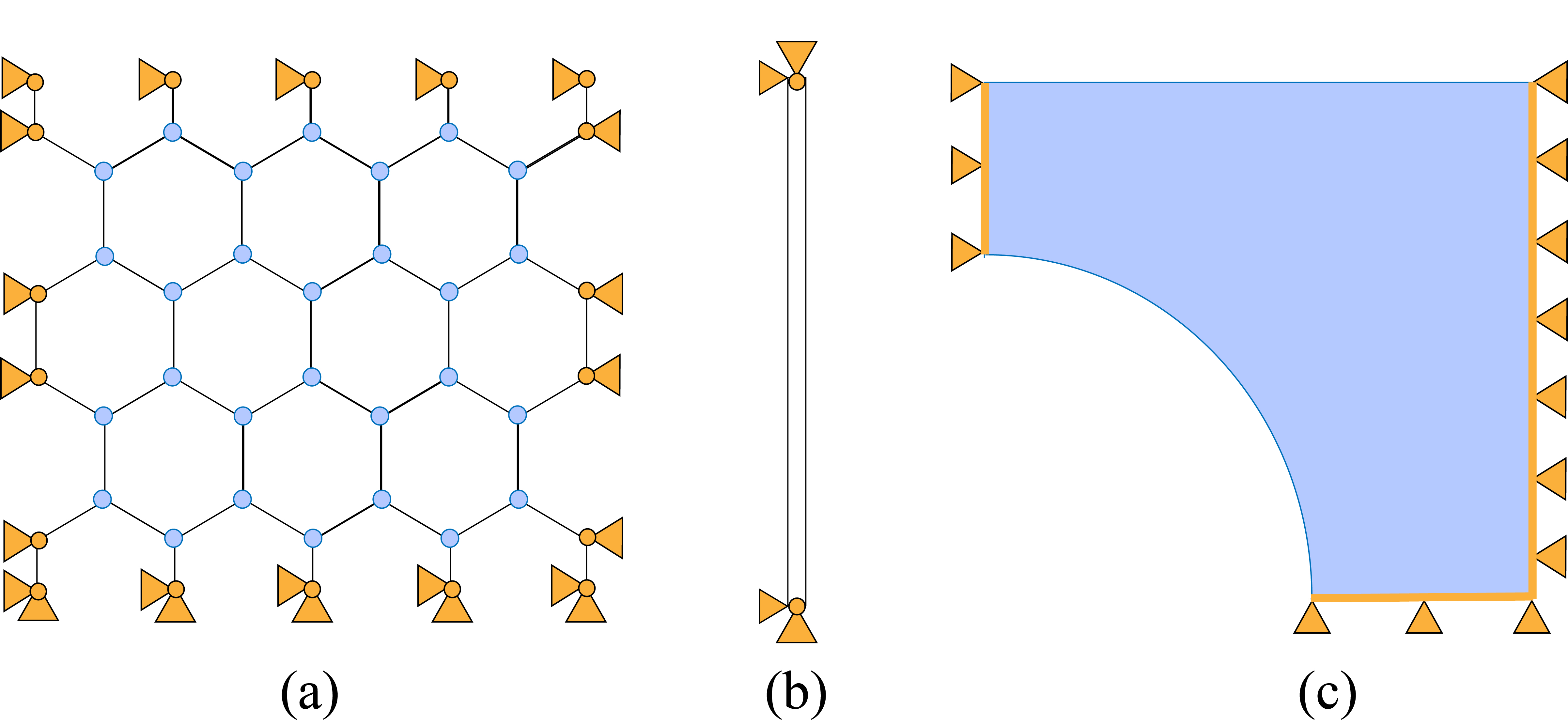}
    \caption{
    \small Schematics of the three distinct case studies: (a) hexagonal truss, (b) Timoshenko beam, and (c) plane-strain continuum. Boundary conditions are prescribed at the locations denoted by yellow dots/lines.}
    \label{fig:sch}
\end{figure}
\section{Example problem cases}
We investigate three different systems subjected to prescribed boundary conditions. In all cases, we use linear elastic, isotropic material properties that are similar to 
steel with density $\rho= 7850 \, \text{kg/m}^3$, Young's modulus $E=21\times 10^4 \, \text{GPa}$, and Poisson's ratio $\nu=0.3$.\\ 

\textbf{Case (I):} The hexagonal truss, as illustrated in Fig.\,\ref{fig:sch}(a), consists of truss members with length $L=1.5 \, \text{mm}$ and circular cross-section radius $r= 0.5 \, \text{mm}$.
Boundary conditions are applied in the following manner: (1) For nodes located at the bottom, displacements in both horizontal and vertical directions are fixed ($u_1=u_2=0$); (2) For nodes at the left, right, and top sides, 
$u_1$ is set to zero, whereas $u_2$ is free. 
In this formulation, a consistent mass matrix (rather than a lumped-mass matrix) is used.\\

\textbf{Case (II): } The Timoshenko beam is illustrated in Fig.\,\ref{fig:sch}(b) with length  $L=9 \,\text{mm}$ and circular cross-section radius $r= 1\,\text{mm}$. 
For boundary conditions, the translational displacements $u_1$ and $u_2$ of both ends are set to zero. Here, we use the lumped-mass formulation for both mass and stiffness matrices.\\

\textbf{Case (III):} The plane-strain continuum model, as depicted in Fig.\,\ref{fig:sch}(c), consists of a quarter section of a square with a circular cutout in its 
geometric 
center. The entire square is  $2\,\text{mm} \times 2\,\text{mm}$, and the open hole has radius $r= 0.5 \,\text{mm}$. 
Boundary conditions are: $x$-symmetric constraint is prescribed on the left edge, while the $y$-symmetric constraint is prescribed on the bottom edge. The right edge is constrained by $u_1=0$.
We also adopt the lumped-mass formulation in this case.

\section{Results and Discussion}
\subsection{Noise-free simulators}
To test the performance, we first execute the FEM-VQE pipeline on a noise-free quantum simulator (e.g., \textit{statevector$\_$simulator} on Qiskit). 
We conduct comprehensive parametric studies to evaluate the 
algorithmic performance in the search for fundamental natural frequency on problems with a wide range of DOFs varying from $8$ to more than $8000$. 
We measure the accuracy of VQE from parametric studies using the percentage error,
\begin{equation}
    Error (\%) = | \lambda_\textrm{q} - \lambda_\textrm{c}|/ \lambda_\textrm{c},
    \label{eq:relative_err}
\end{equation}
where $\lambda_\textrm{q}$ and $\lambda_\textrm{c}$ are eigenvalue estimates from VQE and conventional classical solvers, respectively. We focus on the following three sets of studies using the simulator:

(a) Parametric study on different optimizer types while fixing both the entanglement pattern and depth.

(b) Parametric study on different entanglement patterns while fixing both the optimizer type and entanglement depth.

(c) Parametric study on different entanglement depths while fixing both the optimizer type and entanglement pattern.

\begin{figure}[t!]
    \centering
    \includegraphics[width=1.0\linewidth]{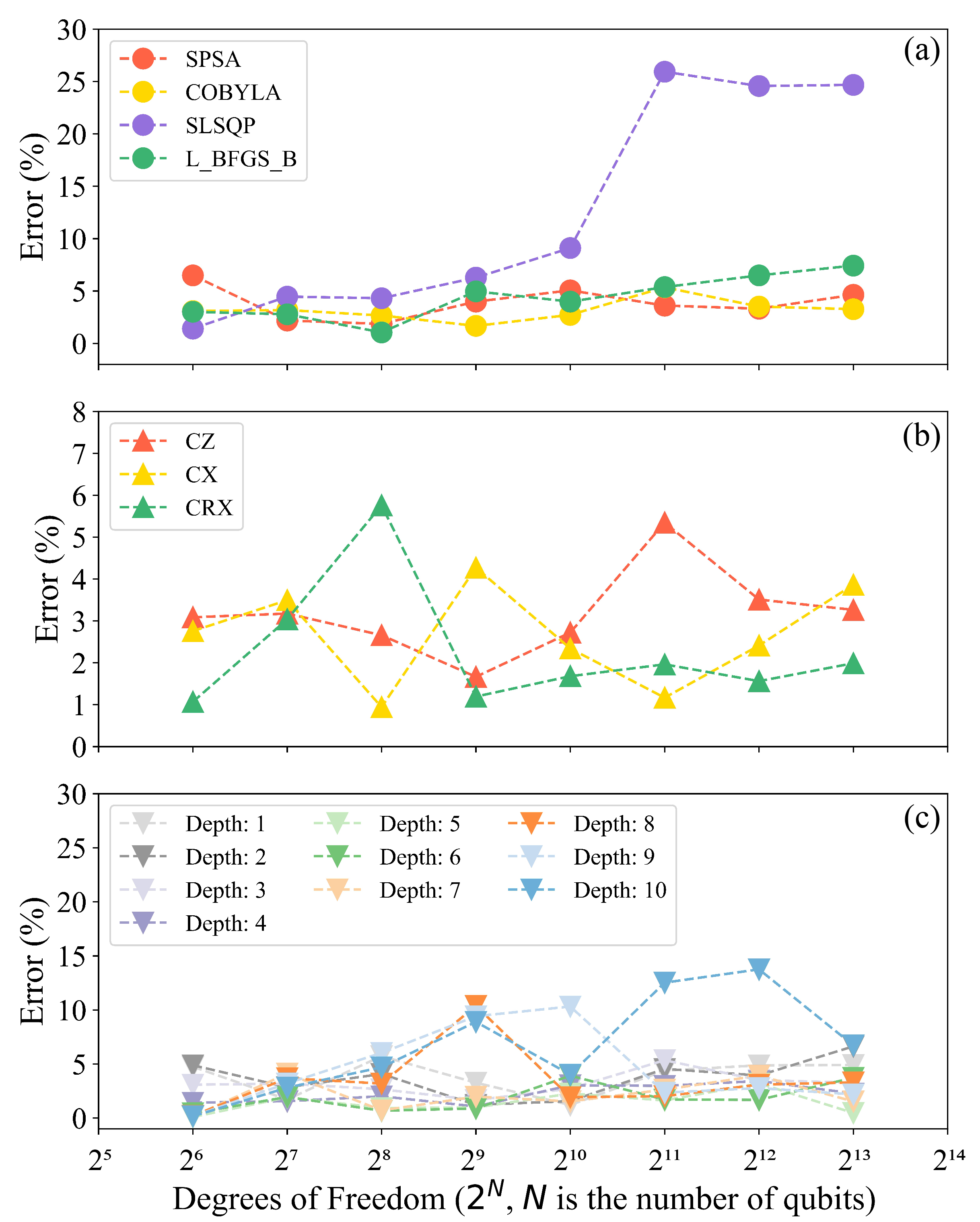}
    \caption{\small Errors of VQE results, as defined by Eq.\,\eqref{eq:relative_err}, for Case (I): (a) Different optimizer choices with entanglement pattern CZ and depth 3. (b) Different entanglement patterns with optimizer COBYLA and entanglement depth 3. (c) Different entanglement depths with optimizer COBYLA and entanglement pattern CZ. 
    }
\label{fig:err_hextruss}
\end{figure}

\begin{figure}[t!]
    \centering
    \includegraphics[width=1.0\linewidth]{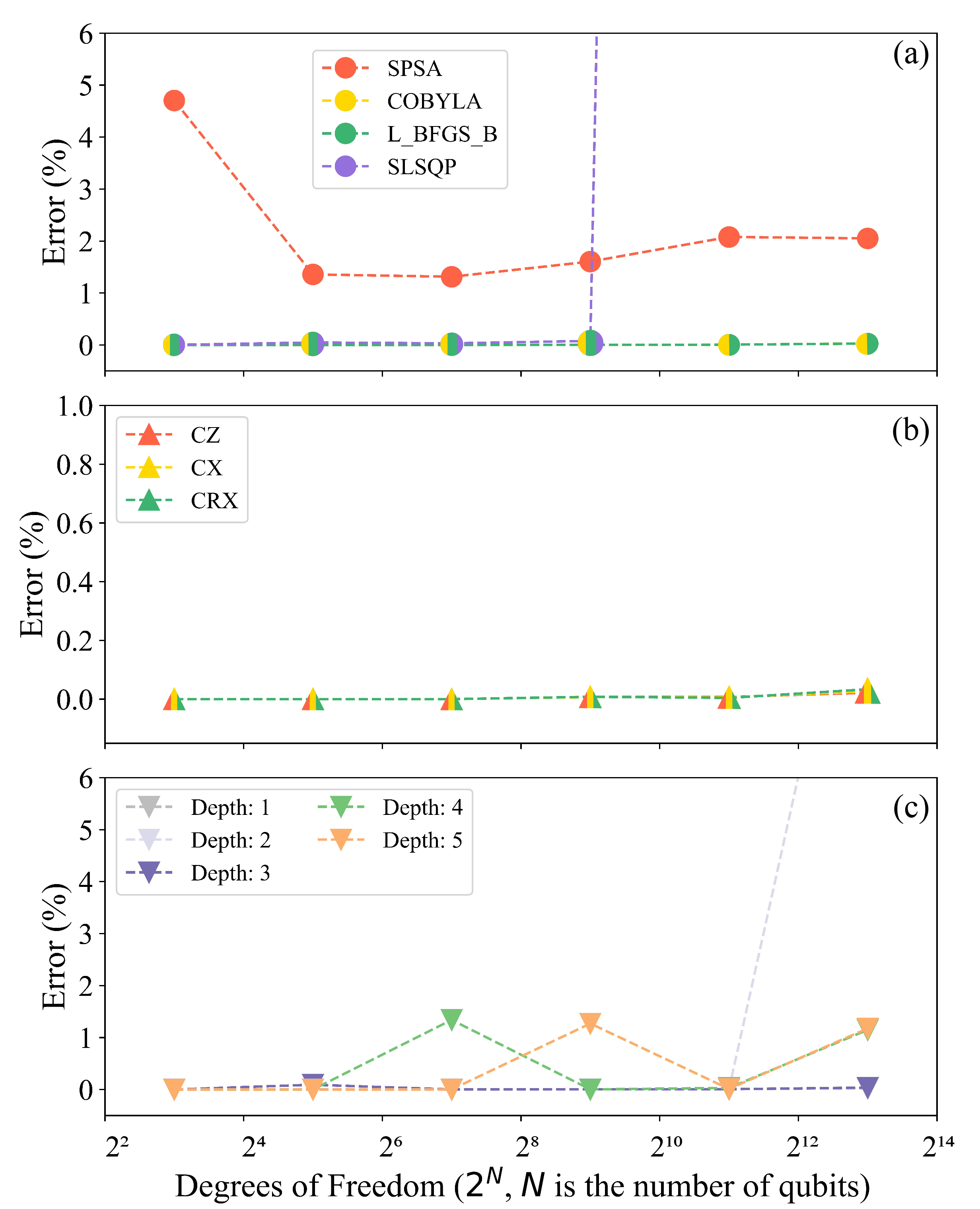}
    \caption{\small 
    Errors of VQE results, as defined by Eq.\,\eqref{eq:relative_err}, for Case (II):
    (a) Different optimizer choices with entanglement pattern CZ and depth 1. Note that, at $N=$ 11 and 13, the errors by SLSQP shoot up to the level of 
    $130\%$, which is out of the range of the plot. 
    (b) Different entanglement patterns with optimizer L-BFGS-B and entanglement depth 1. (c) 
    Different entanglement depths with optimizer L-BFGS-B and entanglement pattern CZ. Note that, at $N = 13$, the error by depth = 2 shoots up to the level of 
    $12\%$, which is out of the range of the plot. 
    }
\label{fig:err_TBeam}
\end{figure}

\begin{figure}[t!]
    \centering
    \includegraphics[width=1.0\linewidth]{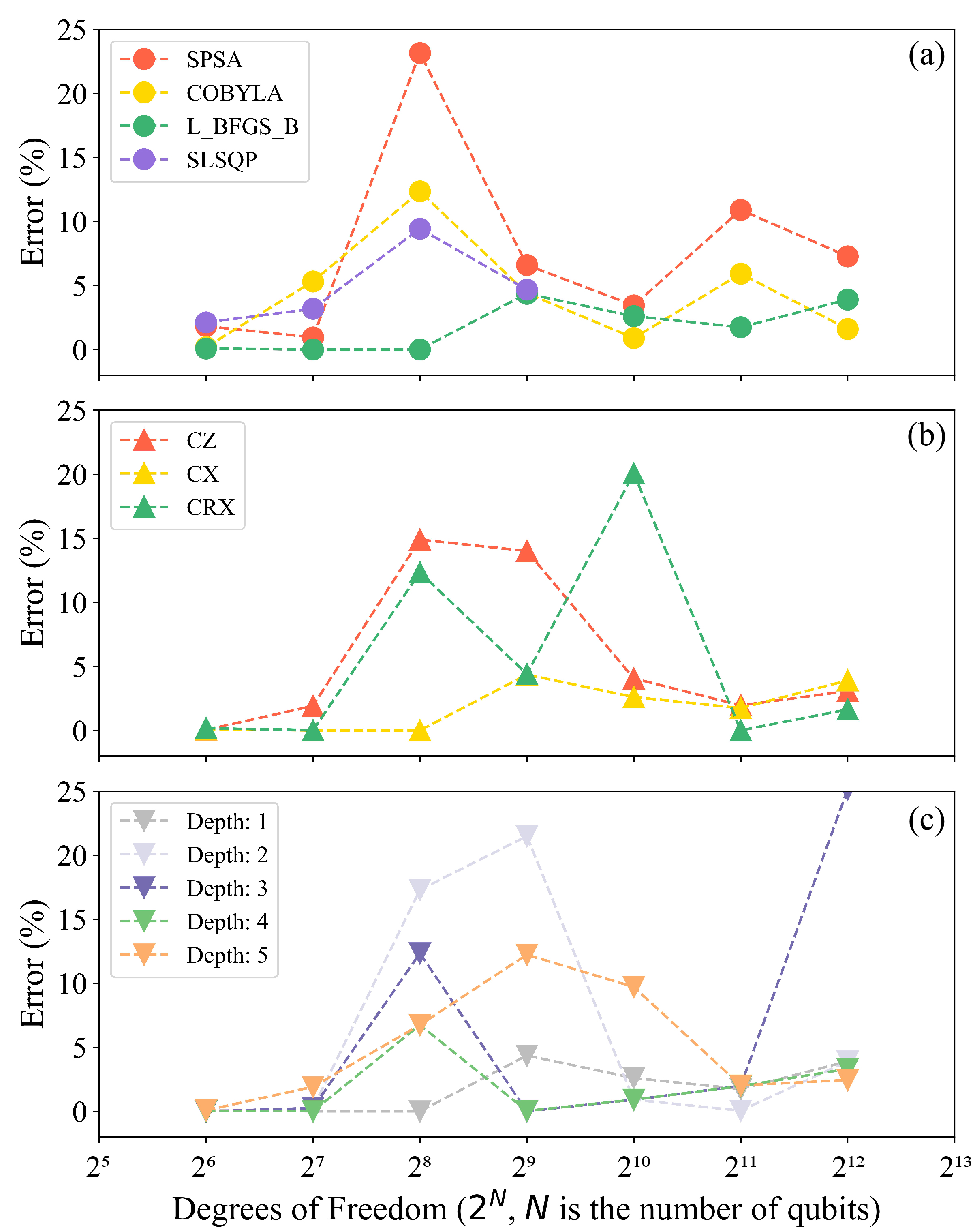}
    \caption{\small  Errors of VQE results, as defined by Eq.\,\eqref{eq:relative_err}, for Case (III):
    (a) Different optimizer choices with entanglement pattern CX and depth 1. (b) Different entanglement patterns with optimizer L-BFGS-B and entanglement depth 1. (c) 
    Different entanglement depths with optimizer L-BFGS-B and entanglement pattern CX. 
    }
\label{fig:err_planestrain}
\end{figure}

We plot the error values defined by Eq.\,\eqref{eq:relative_err} in Figs.\,\ref{fig:err_hextruss}, \ref{fig:err_TBeam}, and \ref{fig:err_planestrain} for all cases. Additional data in terms of convergence rates are presented in the \textit{Supplemental Material}~\cite{SI}.  Our findings reveal that there is no one-size-fits-all particular set of parameters ensuring fast and accurate computation for all three cases. We discuss problem-specific performance characteristics below. We note that all data presented here are reproducible due to the noise-free environment of the simulator. No statistical deviation may occur once all user-defined parameters have been set.\\

Case (I) - 
For the choice of optimizers, Fig.\,\ref{fig:err_hextruss}(a) indicates that, except SLSQP, the error remains under $7.5\%$ for all. There is no discernable difference among the other three optimizers. Likewise, Fig.\,\ref{fig:err_hextruss}(b) does not show any apparent correlation between errors and entanglement patterns. Furthermore, Fig.\,\ref{fig:err_hextruss}(c) shows that the error is always below $5\%$ for any entanglement depth up to 7, and the best performance with errors below $3.5\%$ can be consistently obtained when depth = 4. However, the errors tend to rise above $10\%$ when depth = 8, 9, or 10. Overall, we think an average error of $5\%$ can be expected when applying the VQE algorithm to similar 2D truss problems.\\

Case (II) - 
As illustrated in Fig.\,\ref{fig:err_TBeam}(a), optimizers COBYLA and L-BFGS-B deliver error-free outcomes, while SPSA and SLSQP show inferior performance.
Similarly, Fig.\,\ref{fig:err_TBeam}(b) shows error-free outcomes for all three entanglement patterns, and the largest error observed is only at the level of $0.035\%$. We think this is primarily due to both the lumped-mass-matrix formulation and the high sparsity of the problem-specific Hamiltonian.  
Fig.\,\ref{fig:err_TBeam}(c) displays that, except for depth = 2, the errors remain under $2\%$ for all. Overall, This case study shows that VQE tends to be much more accurate in dealing with similar quasi-1D beam problems.\\

Case (III) - 
Fig.\,\ref{fig:err_planestrain}(a) shows that, among all four optimizer types, only optimizer L-BFGS-B produces errors consistently below $5\%$. Moreover, in contrast to data of Cases (I) and (II), Fig.\,\ref{fig:err_planestrain}(b) seems to show an advantage for using the entanglement pattern CX, which keeps the error under $5\%$. This may be due to the lumped formulation in the setup, which results in a Hamiltonian characterized by independent quantum states and can fully leverage the entanglement. Here, the CX gate enhances correlations by introducing an additional rotation around the X-axis of the Bloch sphere, complementing the single-qubit gates with rotations around the Y- and Z-axes in the ansatz. Although we also adopt the lumped formulation in Case (II), the results among all entanglement types are all negligible there, as shown in Fig.\,\ref{fig:err_TBeam}(b). Hence, no similar advantage of CX is detectable in data from Case (II). In addition, Fig.\,\ref{fig:err_planestrain}(c) supports a definite advantage of shallow circuits, in which the depth = 1 choice keeps the error consistently below $5\%$ in Case (III). Overall, similar to the results in Case (I), we can expect that, through careful choices of the optimizer, entanglement pattern, and entanglement depth, we can expect an average error around $5\%$ of VQE when it is used on plain-strain problems.\\

Furthermore, we also assess the mean of errors (ME) and the standard deviation of errors (SDE) across the different degree-of-freedom (DOF) data points shown in Figs.\,\ref{fig:err_hextruss}, \ref{fig:err_TBeam}, and \ref{fig:err_planestrain}. For instance, in Fig.\,\ref{fig:err_hextruss}(a), employing different types of optimizers 
results in four sets of errors (data illustrated as four different colors), and we calculate the ME and SDE for each optimizer. The assessment of MEs and SDEs on all cases are presented in \textit{Supplemental Material}~\cite{SI}.\\

\subsection{Quantum processing units}
IBM QPUs are hardware designed to execute quantum computations. Employing qubits as their fundamental units of information, they harness the principles of quantum mechanics, which allow them to perform complex calculations far beyond the capability of traditional, binary-bit-based CPUs.

Leveraging 
free and open quantum hardware resources, we execute our FEM-VQE pipeline on 5-qubit (e.g. \textit{ibmq$\_$manila}) and 7-qubit (e.g. \textit{ibm$\_$nairobi}) 
QPUs (e.g. Falcon r5.11H processors) 
via the IBM Qiskit Runtime platform. Guided by the error assessments presented in Figs.\,\ref{fig:err_hextruss}-\ref{fig:err_planestrain}, we choose optimizer COBYLA and entanglement depth 1 in all cases.  
Furthermore, we apply entanglement pattern CZ for Case (I) and (II), and CX for Case (III).

Unlike noise-free simulators, QPUs may suffer statistical deviations 
since random noise could alter the outcome. After conducting multiple (roughly $\sim$10) trials of each calculation on QPUs, we summarize the 
mean VQE estimate $\lambda_\textrm{q}$ and mean error (ME) of each case 
in Table.\,\ref{table:qpu-main}. The data indicate that the algorithm accumulates much more error on QPUs than it does on the noise-free simulator. Even for Case (II), which is the best performer on the simulator, our results show the errors can jump up to more than 30 \% when $N > 5$. Here, the reliability of the FEM-VQE pipeline can be impeded by three principal factors:

First, the QPU capabilities are severely limited by the computational ``quantum volume" ($V_Q$)~\cite{cross2019validating, miller2022improved, kechedzhi2023effective}
available on the IBM Qiskit Runtime platform, which is capable of handling only up to $V_Q = 32$
with a high degree of quality. In contrast, for the entries in Table.\,\ref{table:qpu-main} with $N = 6$ or $N=7$, our VQE ansatz with depth 1 
contain more qubits than the quantum volume can handle simultaneously~\cite{IBM_QV}. 
This limitation leads to larger errors when compared with results from the noise-free simulator, which accommodates up to 5000 qubits without any restrictions on $V_Q$.

Second, transpilation~\cite{earnest2021pulse, younis2022quantum} 
poses additional challenges which inflate circuit metrics including gates, depths, $V_Q$, and error rates, further exacerbating the 
computational efficacy. 
This worsens the QPUs performance since the actual circuit depth can be increased by the transpilation, and deeper circuits produce more errors. In the end, the accumulative error may hinder the update of parameters in the cost function landscape.

Third, the thermal noise and electromagnetic interference on QPUs corrupt the preparation and measurement of quantum state $|\psi(\boldsymbol{\theta}) \rangle$, resulting in 
low-quality of VQE estimates.

Lastly, error accumulation deriving from the variational form of the ansatz and the classical optimizer can impair the accuracy of eigenvalue prediction. The ideal condition would be achieving a fault-tolerant result from an aptly parameterized quantum circuit $U(\boldsymbol{\theta})$ without classical computers, due to the risk of 
a sub-optimal solution caused by a barren plateau problem~\cite{endo2021hybrid, salm2020nisq}. These considerations are inherent and unavoidable aspects of the current technological landscape. Although this also occurs on both the simulator, the effects are more pronounced with QPUs.

\begin{table}[h!]
\setlength{\arrayrulewidth}{0.25mm}
\renewcommand{\arraystretch}{1.5}
\centering
\caption{\centering Errors in QPU computations}
    \begin{tabular}[t]{|p{2.6cm}||p{1.2cm}|p{1.2cm}|p{1.3cm}|}
    \hline
    & $\lambda_\textrm{{c}}$ & $\lambda_\textrm{{q}}$ & ME$(\%)$\\
    \hline
    Case (I) (N=6) & 0.0771 & 0.0855 & 10.895\\
    \hline
    Case (I) (N=7) & 0.0572 & 0.0688 & 20.280\\
    \hline
    Case (II) (N=3) & 0.0343 & 0.0343 & 0.000\\
    \hline
    Case (II) (N=5) & 0.0604 & 0.0606 & 0.331\\
    \hline
    Case (II) (N=7) & 0.0354 & 0.0483 & 36.441\\
    \hline
    Case (III) (N=6) & 0.0457 & 0.0535 & 17.068\\
    \hline
    Case (III) (N=7) & 0.0317 & 0.0380 & 19.874\\
    \hline
    \end{tabular}
\label{table:qpu-main}
\end{table}
\section{Conclusion}
The successful implementation of a FEM-VQE pipeline presented in this Letter is the first step towards harnessing quantum computing to solve problems in solid mechanics and structural engineering. This computational framework could be particularly useful to researchers who wish to take advantage of the noisy intermediate-scale quantum (NISQ) computing devices that are rapidly becoming available now. Our parametric studies on (I) 2D truss, (II) 1D beam, and (III) plane-strain continuum cases provide direct evidence supporting the validity of this quantum-classical hybrid algorithm. 
On a noise-free simulator, our data prescribe the following set of optimal parameters: (I) COBYLA with CZ and depth = 4; (II) L-BFGS-B with CZ and depth = 1;
(III) L-BFGS-B with CX and depth = 1.\\
While the demonstration detailed in this letter does not manifest quantum supremacy over classical computers in terms of accuracy or efficiency, it does validate the integrative methodology that couples \textsc{abaqus} with Qiskit VQE implementation. This methodology holds the potential to address problems encompassing significantly more degrees of freedom as quantum computers continue to become more capable and more widely available. The large error values in our results are not a reflection of the shortcomings of the algorithm. Rather, they are indicative of the current state of quantum computing technology~\cite{IBM_QuantumErrorCorrection, IBM_QuantumRoadMap, IBM_Promise, IBM_127}. The prospect of enhanced quantum processors, fortified with advanced error mitigation techniques and designed to operate at a utility-scale with improved quality, is an anticipated advancement that warrants keen attention~\cite{kim2023evidence,van2023probabilistic}. For example, quantum hardware manufacturers are now making devices 
with up to 27, 65, 127, and 433 qubits available to the general research community~\cite{IBM_Newsroom}. This could potentially enable our FEM-VQE pipeline to solve mechanics problems with $10^{8}$, $10^{19}$, $10^{38}$, and $ 10^{130}$ DOFs.


\section{Acknowledgement}
YL and PW are supported by both the Research Incentive Seed Grant Program and the start-up research funds of the Department of Mechanical Engineering at the University of Utah. 
JR and JL gratefully acknowledge support via National Science Foundation (NSF) FM 2036881. JL was also partially supported by the NSF MRSEC program under award DMR-1720530.
The support and resources from the Center for High-Performance Computing at the University of Utah are gratefully acknowledged. 

We acknowledge the use of IBM Quantum services for this work. The views expressed are those of the authors and do not reflect the official policy or position of IBM or the IBM Quantum team. In the process of developing the code necessary for this project, we also immensely benefited from the resources and community support provided by the IBM Qiskit channel. The coding techniques and strategies derived from this resource were instrumental in overcoming computational challenges and refining the efficiency of our algorithms. We wish to express our gratitude for valuable assistance and guidance from the Qiskit community.

The authors declare no business or financial connections with either ABAQUS Inc. or IBM Corp. 
\normalem
\bibliography{references}

\end{document}